\documentclass[twocolumn,pra,aps,showpacs,nofootinbib,nobalancelastpage]{revtex4}
\usepackage{graphicx}

\newcommand{\smallfrac}[2]{\mbox{$\frac{#1}{#2}$}}

\begin{document}


\title{How to project qubits faster using quantum feedback
      }

\author{Kurt Jacobs}
\affiliation{65 Abraham heights, Nelson 7001, Auckland, NZ}

\begin{abstract}
When one performs a continuous measurement, whether on a classical or quantum system, the measurement provides a certain average rate at which one becomes certain about the state of the system. For a quantum system this is an average rate at which the system is projected onto a pure state. We show that for a standard kind of continuous measurement, for a qubit this rate may be increased by applying unitary operations during the measurement (that is, by using Hamiltonian feedback), in contrast to the equivalent measurement on a classical bit, where reversible operations cannot be used to enhance the rate of entropy reduction. We determine the optimal feedback algorithm and discuss the Hamiltonian resources required.
\end{abstract}


\maketitle

It was discovered recently (\cite{DJJ,FJ}) that the average amount by which the quantum state of a system is purified during a measurement depends, in general, on the basis in which the measurement is made. Since changing the basis of a measurement is equivalent to performing a unitary transformation on the system, we can re-state this property by saying that a unitary transformation may be applied to a system, so as to increase the average amount of information that the measurement provides about the final (post-measurement) state. This restatement makes it particularly clear what is being changed about the measurement in order to enhance the information: merely the addition of a separate Hamiltonian evolution. Taking this point of view, the ability to perform unitary transformations, or equivalently, Hamiltonian evolution, can be considered as a resource which can be used to enhance the properties of a fixed measurement process. 


Consider now a continuous measurement of, for example, a qubit. This may be represented by a sequence of identical `finite strength' measurements, each of which partially projects the qubit onto the basis $\{|0\rangle,|1\rangle\}$~\cite{DJJ} (We will refer to this as the {\em computational basis}, by which is simply meant the basis in which information is encoded). As more measurements are made, the purity of the state of the qubit increases, until eventually the qubit ends up in one of the basis states. Examining this process, we find that the state that results from a measurement in the sequence is not ideal for the purposes of purification for the next measurement in the sequence. It is therefore possible to perform a unitary transformation at the end of each measurement (and depending on the measurement result) to increase the average rate at which the state is purified. In the continuum limit this becomes a continuous feedback process.

This fact, while interesting as it provides a technique for enhancing a continuous measurement, is, if anything, more interesting from a fundamental point of view, due to the fact that the same measurement process performed on a classical bit cannot be enhanced by a reversible transformation, since the necessary sequence of transformations requires that superposition states of the computational basis must be available. Thus, this constitutes an example of something which the quantum nature of an object makes possible. 

Since this feedback algorithm is designed to enhance the properties of the measurement itself, the resulting measurement process is an example of an {\em adaptive} measurement, a term first introduced, we believe, by Wiseman~\cite{Wadapt}. Wiseman's adaptive measurement scheme, realized recently by Armen {\em et.\ al.}~\cite{Armen}, was designed to make a canonical phase measurement optimally well, given that, in practice, one only has arbitrary quadrature measurements at ones disposal. The topic discussed here is therefore a different kind of application for adaptive measurements.

Before we begin the development of the feedback algorithm, let us say a few words about quantum and classical measurements, and the relationship between them. Classical measurements on classical systems are described by Bayesian inference, and may be written in the same form as quantum measurements. If one writes the classical probability distribution of the quantity one is measuring as a diagonal density matrix, then classical measurements are in fact a subset of quantum measurements: While quantum measurements are described by a set of operators $\Omega_n$, where the only restriction is that $\sum_n\Omega_n^\dagger\Omega_n = 1$, classical measurements have the further restriction that all the $\Omega_n$ commute with the density matrix describing the classical system (for a fuller discussion see~\cite{KJ}). 


The behavior of a quantum measurement with commuting operators therefore reduces to that of a classical measurement when either unitary transformations are not available to rotate the qubit out of the computational basis, or there is a rapid decoherence process which immediately decoheres the qubit in the computational basis. In that case the unitary transformation becomes merely a classical diffusion process acting on the bit.


Here we will be interested in a widely applicable model of continuous measurement, where the measurement record is a Wiener process. We will measure in the z-basis of a spin-half system (that is, measure the observable $\sigma_z$), denoting the basis states as $\{|\pm\rangle\}$ (thus the final result of the measurement will either be $|-\rangle$ or $|+\rangle$). (Any further reference to `the computational basis' will refer to these states.) To describe the continuous measurement one applies a POVM in each small time interval $\Delta t$, where the POVM is chosen to scale with time in such a way that a sensible continuum limit exists~\cite{CM}. The resulting continuous measurement is not merely a mathematical curiosity, as it corresponds to real measurements on physical systems~\cite{PhysMeasGen,PMQbit1,PMQbit2,PMQbit3,PMQbit4} (in particular, Korotkov (\cite{PMQbit1,PMQbit2,PMQbit3,PMQbit4}) gives explicit examples for solid-state qubits). While the approach in~\cite{CM} uses a POVM in each time step which has an infinite number of outcomes, for measurements on a two-state system one can alternatively employ a POVM with two-outcomes. A two-outcome measurement which provides information about the computational basis is
\begin{eqnarray}
 \Omega_\pm & = & \sqrt{\kappa}|\pm\rangle\langle\pm| + \sqrt{1-\kappa}|\mp\rangle\langle\mp| \nonumber \\
          & = & \frac{1}{2}\left[ (\sqrt{\kappa}+\sqrt{1-\kappa}) I \pm
	       (\sqrt{\kappa}-\sqrt{1-\kappa}) \sigma_z\right],
\end{eqnarray}
We will refer to this measurement in what follows as ${\cal M}$.
When $\kappa = 1/2$ ${\cal M}$ provides no information about the quantum system, leaving the state of knowledge unchanged, and when $\kappa = 0$ or $1$ the operators $\Omega_\pm$ are rank one projectors, so that ${\cal M}$ provides the maximal amount of information about the final state; this case is an `infinite strength' measurement by the terminology of~\cite{DJJ,FJ}.

Setting $\kappa = 1/2 - \sqrt{2\gamma\Delta t}$, and taking the limit of repeated measurements as $\Delta t \rightarrow 0$, the evolution of the density matrix describing the observers state of knowledge, $\rho$, is given by the stochastic Schr\"odinger equation 
\begin{equation}
  d\rho = -\gamma [\sigma_z,[\sigma_z,\rho]] dt + \sqrt{2\gamma} (\{ \sigma_z, \rho \}_+ - 2\langle\sigma_z\rangle) dW ,
\end{equation}
where $dW$ is the Gaussian stochastic Wiener increment satisfying $dW^2 = dt$. 

We are interested here in how the observer's uncertainty of the quantum state reduces over time. In order to keep the calculations tractable, we will use as our measure of uncertainty the so called ``linear entropy'', $P(\rho) = 1 - \mbox{Tr}[\rho^2]$. (That this is a useful measure of uncertainty is due to its concavity~\cite{FJ}.) For a completely mixed state, $P(\rho) = 1/2$, and for a pure state, $P(\rho) = 0$. 

In general, the reduction in the linear Entropy will depend upon the outcome of the measurement. As such, a sensible measure of the purifying power of the measurement is therefore the {\em average} reduction in the linear entropy, over the two outcomes. From~\cite{FJ} we know that the amount by which, on average, the two-outcome measurement purifies the state (that is, reduces $P(\rho)$) is 
\begin{eqnarray}
  \Delta_P & = & (2\kappa -1) P \frac{1 - (1-2P)\cos^2\theta}
                                    {1 - (1-2P)(2\kappa -1)^2\cos^2\theta} ,
\end{eqnarray}
where we have written the initial density matrix as $ \rho = (I + {\bf a}\cdot{\bf \sigma})/2$, with $\theta$ being the angle between ${\bf a}$ and the $z$-axis, and $|{\bf a}|^2 = (1-2P)$. The reduction in uncertainty is maximized when $\theta=\pi/2$. That is, when the basis in which the density matrix is diagonal is maximally different from the $z$-basis, being the basis in which the measurement is made, the measurement is most effective in purifying the state. This somewhat curious result means that, if the observer's state-of knowledge is not maximally complementary to the measurement basis, and the observer's objective is purification, then a unitary transformation should be applied to the state prior to the measurement. It is also worth noting that, when the average reduction in the entropy is greatest, this reduction is the same for both measurement outcomes. Thus, in this case, the entropy reduction is deterministic, not random.


Now let us consider the consequences of this for a continuous measurement, performed on an initially completely mixed state. Approximating the continuous measurement by a sequence of measurements ${\cal M}$, we see that when we make the first measurement, $\Delta_P$ is maximal. However, the result of the first measurement does not produce a state diagonal in a basis complementary to the $z$ basis, and this is generally true of the state resulting from a measurement in the $z$-basis. As a result, the average purification will not be optimal for at least the majority of the measurements ${\cal M}$, and thus for essentially all of the duration of the continuous measurement process. 

This suggests, therefore, the following procedure: After each measurement we apply a unitary transformation to rotate the state appropriately, so as to achieve maximal $\Delta_P$ for each measurement in the sequence (each measurement step). Taking the continuum limit, this results in a continuous feedback algorithm which increases the rate of projection during the continuous measurement. Calculating the behavior of the linear entropy, $P$, as a function of time is straightforward, because, as mentioned above, when $\Delta_P$ is maximal, the change in $P$ is the same for both outcomes of the measurement. Hence, for a finite sequence of steps we have, for the $n^{th}$ step, 
\begin{eqnarray}
  P_n & = & \left( 1 - b^2/n \right)^n P_0 ,
  \label{Pn}
\end{eqnarray}
where $b^2 = (2\kappa -1)$. In the continuum limit this becomes 
\begin{eqnarray}
  P(t) & = & e^{-8\gamma t} P(0) .
\end{eqnarray}
Without the feedback process, the evolution of $P(t)$ depends on the measurement outcomes, and as a result is stochastic. Thus, not only does the Hamiltonian feedback algorithm increase the rate at which the state is purified, but also changes the entropy from being a random function of time to a deterministic one. (In fact, it is clear that this is true for any entropy, linear, von Neumann, or otherwise.)

\begin{figure}
\leavevmode\rotatebox{-90}{\includegraphics[width=0.7\hsize]{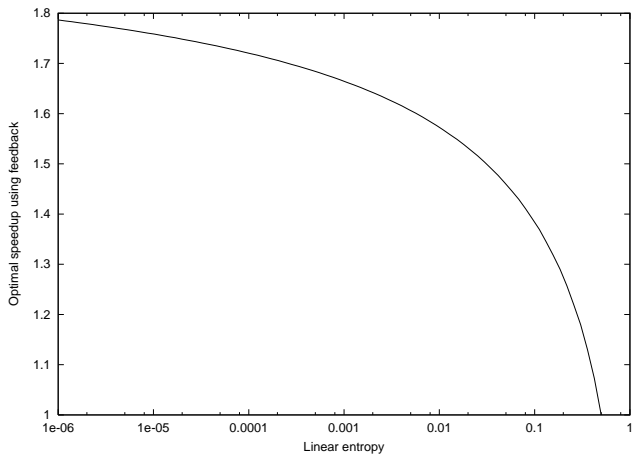}}
\caption{The speed-up factor provided by the optimal Hamiltonian feedback algorithm, as a function of final purity, for a continuous measurement in a given basis of a qubit. The case displayed is for an initially completely mixed state.}
\label{fig1}
\end{figure}

In the absence of feedback, due to its stochastic nature, the behavior of $P(t)$ is much harder to obtain, even for the simple canonical continuous (classical!) measurement we have here. Using the technique in~\cite{JK}, one obtains, for an initially completely mixed state,
\begin{eqnarray}
  P_{\mbox{\scriptsize c}}(t) & = & \frac{e^{-4\gamma t}}{\sqrt{8\pi t}} \int_{-\infty}^{+\infty} \frac{e^{-x^2/(2t)}}{\cosh(\sqrt{8k}x)} dx ,
  \label{Pclass}
\end{eqnarray}
where we use the subscript $c$ to denote the fact that this is also the result for a classical continuous measurement on a classical bit. 
An analytic solution for the integral in Eq.(\ref{Pclass}) does not appear to exist, and we therefore evaluate it numerically. In figure~\ref{fig1} we plot the speed-up factor provided by the feedback algorithm in obtaining a given final level of purity, when the initial state is completely mixed. This factor is independent of $\gamma$, and increases with the final purity, tending to 2 as the final entropy tends to zero.

The Hamiltonian feedback algorithm above has been obtained simply by optimizing the increase in purity of the final state at each measurement step. We now show that this is, in fact, the optimal feedback algorithm for obtaining the maximum purity of the state at any particular future time. 
First, let us denote the map which takes us from an initial linear entropy $P$, to the final linear entropy $P'$, for a single time step, when we use the optimal unitary transformation, as $f(P)$. Thus we have $P'=f(P)=(1-b^2)P$, so $f$ is linear in $P$. 


Now, let us consider a single measurement step ${\cal M}$, where the initial entropy is $P$. If we use a sub-optimal procedure, then one of two states results, and we can label the linear entropy of these as $P_1 = g_1(P)$ and $P_2 = g_2(P)$, respectively. On the other hand, the entropy that results from the optimal procedure is $f(P)$, and the three entropies satisfy
\begin{eqnarray}
  \sum_i^n p_i P_i & \ge & f(P)  ,   \label{ineqP}
\end{eqnarray}
where $p_1$ and $p_2$ are the respective probabilities that the two sub-optimal entropies were obtained. If we apply the optimal procedure to both results (both sides) then since $f(P)$ is linear, we have 
\begin{eqnarray}
  \sum_{i=1}^2 p_i f(g_i(P)) & \ge & f(f(P)) .          
\end{eqnarray}
Alternatively, if we were to apply a non-optimal procedure to both cases (both sides of Eq.(\ref{ineqP})), then we would have, for each of the 
outcomes $i=1,2$,
\begin{eqnarray}
  \sum_{j=1}^2 p^i_j g^i_j(g_i(P))  & \ge &  f(g_i(P)) .  \label{ineqP2}
\end{eqnarray}
Thus, by the end of the second step, two sequential non-optimal procedures would give
\begin{eqnarray}
  \sum_{i=1}^2 p_i \left[ \sum_{j=1}^2 p^i_j g^i_j(g_i(P) \right] 
                    & \ge & \sum_{i=1}^2 p_i f(g_i(P)) \nonumber \\
                    & \ge & f(f(P))
\end{eqnarray}
where the first line uses Eq.(\ref{ineqP2}), and the second uses the linearity of $f(P)$. Thus, after two steps the result of non-optimal measurements in both steps gives an average entropy which is higher than the result of two optimal steps. Clearly this procedure can be repeated $n$ times to obtain the result for $n$ measurement steps. Thus, we can write the final result of $n$ non-optimal steps as  
\begin{eqnarray}
  \sum_{i=1}^{2^n} p_i P_i & \ge & f^{\mbox{\scriptsize o} (n)} (P)             \label{ineqAll}
\end{eqnarray}
for some $p_i$ and $P_i$. In order for the procedure which uses all optimal steps to render a greater entropy than a procedure that uses some non-optimal steps, then it would have to be possible to apply an optimal step to the left hand side of Eq.(\ref{ineqAll}), such that $\sum_{i=1}^2 p_i f(P_i) < f^{\mbox{\scriptsize o} (n+1)} (P)$.
However, since $f(P)$ is linear, this is impossible. We can therefore conclude that the Hamiltonian feedback algorithm presented above gives the maximum possible entropy reduction for any number of steps, or, in the continuum limit, the maximum possible entropy reduction for a measurement of any given duration.


So far we have discussed the effects of the feedback algorithm, but not given explicitly the algorithm itself. The unitary transformation required after each measurement step must be such so as to rotate the qubit so that the Bloch vector lies in the $xy$-plane. The first thing to note is that the minimum angle of rotation required to do this is achieved by rotating such that the $x$ and $y$ elements of the Bloch vector remain in the same proportions (i.e. by keeping the angle that the Bloch vector makes with the $x$ and $y$ axes the same). The second is that an application of the measurement ${\cal M}$ also keeps these angles the same. Assuming that the initial state is either completely mixed, or has been rotated prior to the measurement so that the Bloch vector lies in the $xy$-plane (which is the optimal thing to do), then the initial $a_z=0$, and the state after a measurement ${\cal M}$ is 
\begin{eqnarray}
  \rho_f = \smallfrac{1}{2}(I + \sqrt{1-b^2}(a_x \sigma_x + a_y \sigma_y) \pm b \sigma_z) .
\end{eqnarray}
The unitary transformation required to rotate such a state so that once again $a_z =0$, using the minimum angle of rotation, is $U = \exp[-i(\alpha/2){\bf n}\cdot{\bf \sigma}]$, with
 \begin{eqnarray} 
 \alpha & = & \mbox{tan}^{-1}\left[ \pm b/\left( \sqrt{1-2P}\sqrt{1-b^2} \right) \right] , \\
 {\bf n} & = & (-\cos(\phi),\sin(\phi),0) ,
\end{eqnarray}
and where we have used the fact that $ |{\bf a}|^2 = a_x^2 + a_y^2=1-2P$, and defined $\phi$ by $a_x = |{\bf a}|\cos(\phi)$. 
Since $\phi$ remains the same throughout the sequence of measurements, ${\bf n}$ remains unchanged throughout the feedback process, and it is merely $\alpha$ that changes at each feedback step. After the $n^{th}$ measurement $P_n$ is given by Eq.(\ref{Pn}), and hence 
\begin{eqnarray} 
  \alpha_n & = & \mbox{tan}^{-1} \left[\pm b / ( \sqrt{1-2P_0(1-b^2/n)^n}\sqrt{1-b^2} ) \right] . \nonumber
\end{eqnarray}
Note that, when the initial state is completely mixed, the rotation angle $\alpha$, after the first measurement is always $\pi/2$, regardless of the strength of ${\cal M}$ (i.e. regardless of the value of $b$).

In the continuum limit, the feedback angle $\alpha(t)$ becomes
\begin{eqnarray} 
  \alpha(t) & = & \mbox{tan}^{-1} \left( \frac{\sqrt{8\gamma} dW} 
                        {\sqrt{1-2P(0)e^{-8\gamma t}}} \right)   \nonumber \\
	    & = & \frac{\sqrt{8\gamma} dW}{\sqrt{1-2P(0)e^{-8\gamma t}}}, \;\;\;
	          P(t) < 1/2 .
\end{eqnarray}
Since the Hamiltonian required to rotate through $\alpha$ is proportional to $\alpha(t)/dt$, the Hamiltonian required for the feedback is proportional to the measurement noise $dW/dt$. Thus the feedback required to obtain an optimal projection rate is Wiseman-Milburn type Markovian feedback, with the addition of a time dependent factor. However, this kind of feedback is strictly speaking an idealization valid in the limit of a large Hamiltonian, since $dW/dt$ is infinite. In addition, the feedback Hamiltonian diverges for an initially completely mixed state, since $\alpha(0)=\pi/2$. Thus, a real continuous feedback procedure will provide a lower rate of projection, depending on the available Hamiltonian resources. We note that the divergence of the feedback Hamiltonian for $P=1/2$ has an analogue in Wiseman's adaptive phase measurement; in that scheme, for this state, it is the rate of change of the phase estimate which diverges. It is interesting that for both schemes this divergence is associated with the fact that the measurement must break the symmetry of the state.


While the feedback algorithm projects a qubit onto a final pure state with maximal speed, it is fairly clear from the construction of the process that the operations corresponding to the many possible final outcomes will be mutually non-orthogonal. As a result, this adaptive measurement will almost certainly not provide full information regarding the initial preparation of the qubit.
This is why we refer to the algorithm as projecting, rather than measuring the qubit, for the sense of the term ``measurement'' contains a certain ambiguity: it could mean either obtaining information about the initial preparation, useful in classical and quantum communication, or the final state, useful in quantum feedback control and quantum state preparation. While in this case the quantum state is projected maximally fast, this is at the expense of losing information about the initial preparation. This raises the question of whether there is a trade-off between speed of projection, and loss of initial information, in the kind of measurements considered here. In addition, optimal algorithms for projecting higher dimensional systems, and optimal rates obtainable with fixed Hamiltonian resources, are also open questions for further work.

The author would like to thank Howard Wiseman for helpful comments on the manuscript.    



\end{document}